\documentclass[%
reprint,
superscriptaddress,
amsmath,amssymb,
aps,
]{revtex4-1}

\usepackage{graphicx}
\usepackage{bm}
\usepackage[dvipsnames]{xcolor}
\usepackage{tabularx}
\newcolumntype{Y}{>{\centering\arraybackslash}X}

\newcommand{\atan}{\mathrm{atan}}

\newlength{\figwidth}
\newcommand\trick[1]{}
\begin{document}

\title{Lossy Quantum Defect Theory of Ultracold Molecular Collisions}
\author{Arthur Christianen}
\affiliation{Max-Planck-Institut f{\"u}r Quantenoptik, 85748 Garching, Germany}
\affiliation{Munich Center for Quantum Science and Technology, 80799 M{\"u}nchen, Germany}
\author{Gerrit C. Groenenboom}
\affiliation{Radboud University, Institute for Molecules and Materials, Heijendaalseweg 135, 6525 AJ Nijmegen, the Netherlands}
\author{Tijs Karman}
\email{t.karman@science.ru.nl}
\affiliation{Radboud University, Institute for Molecules and Materials, Heijendaalseweg 135, 6525 AJ Nijmegen, the Netherlands}
\date{\today}
\begin{abstract}
We consider losses in collisions of ultracold molecules described by a simple statistical short-range model that explicitly accounts for the limited lifetime of classically chaotic collision complexes.
This confirms that thermally sampling many isolated resonances leads to a loss cross section equal to the elastic cross section derived by Mayle \emph{et al}. [Phys.\ Rev.\ A {\bf 85}, 062712 (2012)],
and this makes precise the conditions under which this is the case.
Surprisingly, we find that the loss is nonuniversal.
We also consider the case that loss broadens the short-range resonances to the point that they become overlapping.
The overlapping resonances can be treated statistically even if the resonances are sparse compared to $k_BT$, which may be the case for many molecules.
The overlap results in Ericson fluctuations which yield a nonuniversal short-range boundary condition that is independent of energy over a range much wider than is sampled thermally.
Deviations of experimental loss rates from the present theory beyond statistical fluctuations and the dependence on a background phase shift are interpreted as non-chaotic dynamics of short-range collision complexes.
\end{abstract}

\maketitle

\section{Introduction}

Ultracold molecules are promising for applications in quantum simulation \cite{micheli:06, buchler:07,pupillo:08, krems:09,yan:13}, quantum computing \cite{demille:02,yelin:06,park:17,ni:18} and precision measurement \cite{carr:09, krems:09,andreev:18}.
However, these applications are held back by rapid collisional loss \cite{ni:08,danzl:10,takekoshi:14,molony:14,park:15,guo:16,rvachov:17,seesselberg:18,yang:19}.
In the language of quantum defect theory (QDT) \cite{mies:84a,mies:84b,burke:98}, short-range loss is described by a parameter $y$ between $0$ and $1$,
which indicate no short-range loss and complete short-range loss, respectively -- also known as universal loss \cite{idziaszek:10}.
The probability of short-range loss is related to this parameter as $4y/(1+y)^2$.
Collisional losses of KRb \cite{ni:08}, NaK \cite{park:15,yan:20,bause:21,gersema:21}, NaRb \cite{ye:18,guo:18,gersema:21}, and CaF \cite{cheuk:20} molecules are consistent with universal loss,
although careful measurements of loss in RbCs+RbCs collisions could resolve nonuniversal behavior and pinpoint the loss parameter $y=0.26(3)$ \cite{gregory:19}.
While nonuniversal, this still corresponds to a 65~\% probability of loss in short-range encounters between RbCs molecules,
which is surprisingly high and deserving of an explanation as these molecules are nonreactive.

Mayle \emph{et al.}\cite{mayle:12,mayle:13} proposed the mechanism of ``sticky collisions'' to describe loss of nonreactive molecules.
The idea is that collision complexes consisting of two ultracold molecules may have a very long lifetime due to their high density of states.
It is believed that this high density of states results in classically chaotic dynamics of the collision complexes,
as supported by classical simulations~\cite{croft:14,klos:21}, quantum reactive scattering calculations~\cite{croft:17,croft:17b}, and experimental product distributions~\cite{liu:21}.
If these lifetimes are long enough for the collision complexes to undergo collisions with another molecule, even nonreactive molecules could be lost by three-body recombination.
Subsequently, Christianen \emph{et al.}\cite{christianen:19b} showed that the lifetimes of these collision complexes were overestimated and in fact are not long enough for three-body recombination to explain the observed close-to-universal loss.
Instead, it was shown that complexes of bialkali molecules can be efficiently excited by the trapping laser \cite{christianen:19a}.
Under typical experimental conditions, the excitation rate of collision complexes is fast compared to the dissociation rate of the collision complexes,
such that essentially any complex formed would be lost by excitation, offering an explanation of the nearly universal loss of nonreactive molecules.
These predictions were later verified by experiments on KRb\cite{liu:20} and RbCs\cite{gregory:20} molecules,
in which complex lifetimes and excitation rates were measured in intensity modulated dipole traps.

\begin{figure}
\begin{center}
\includegraphics[width=\figwidth]{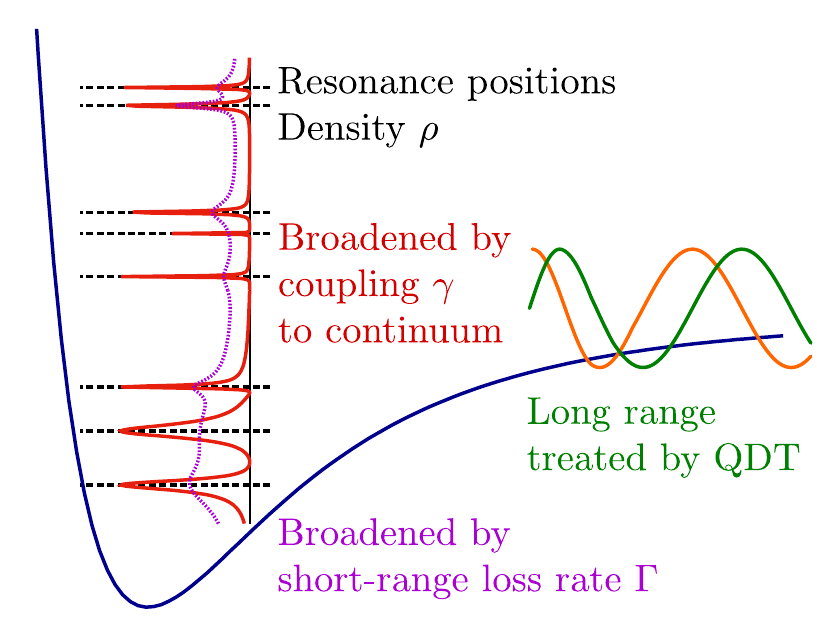}
\caption{ \label{fig:cartoon}
{\bf Illustration of the model considered here.}
The short-range physics is dominated by a set of classically chaotic resonance states, with density $\rho$, described statistically.
The short-range states are broadened by coupling with strength $\gamma$ to the continuum, which is described by quantum defect theory (QDT).
Furthermore, the short-range states undergo losses described by a decay rate $\Gamma$, which leads to further intrinsic broadening.
}
\end{center}
\end{figure}

Apart from the physical origin of the short-range loss of nonreactive molecules,
there are still many open questions surrounding original ideas of sticky collisions.
In particular, Mayle \emph{et al.}\cite{mayle:12,mayle:13} computed \emph{elastic} cross sections thermally averaged over many isolated resonances
and claimed these coincide with the cross section for \emph{universal loss},
and associate this cross section with complex formation.
However, it is unclear how the cross sections are modified if short-range loss is incomplete, at what point this occurs, and how one can interpret observed deviations from universal loss.
More importantly, the revised density of states reported by Christianen \emph{et al.}\cite{christianen:19b} indicate that at microkelvin temperatures RbCs may be the only polar bialkali for which one should expect to thermally sample several resonances,
and the validity of the central assumption of Mayle \emph{et al.}, that one statistically samples many resonances, is questionable.

Here, we present an extended statistical model that accounts for the limited lifetime of the resonance states, as illustrated in Fig.~\ref{fig:cartoon}.
We consider various limiting cases for the short-range loss rate,
summarized in Table~\ref{table}.
For isolated resonances, we confirm that the loss cross section coincides with the elastic cross section derived by Mayle \emph{et al.}\cite{mayle:12,mayle:13},
and quantify how rapid short-range loss needs to be in order for this to hold.
However, we find that the resulting loss cross section is nonuniversal.
Rapid loss broadens the short-range resonances such that these overlap and many resonances contribute at any collision energy.
This enables a statistical treatment even if the density of resonances is low compared to the range of energies sampled thermally.
The resulting loss is substantial but nonuniversal even if short-range loss is fast compared to dissociation.

\begin{table*}
\begin{center}
\caption{ \label{table}
Summary of the different limiting cases considered in this work, the predictions for the cross section or loss parameter, $y$, and the energy dependence.
}
\begin{tabular}{llll}
\hline\hline
Cases considered & & Prediction & Energy dependence \\
\hline
Lossless resonances & $\Gamma = 0$ &   \begin{tabular}{@{}l@{}}Elastic cross section $\sigma^\mathrm{el} = \sigma^\mathrm{univ} / 4(1+\tan^2\lambda)$ \\ See Eq.~\eqref{eq:mayle_el}.\end{tabular} & \begin{tabular}{@{}l@{}}Mean behavior when thermally\\averaged over many\\ resonances\\ $\rho k_BT \gg 1$\end{tabular} \\
\\
Lossy resonances \vspace{4pt} & $\Gamma>0$ & & \\
\hspace{10pt} Isolated resonances & $\rho\bar\Gamma < 1$ & & \\
\hspace{20pt} Fast short-range loss & $\bar\Gamma \gg \bar\gamma C^{-2}$ & \begin{tabular}{@{}l@{}}Loss cross section $\sigma^\mathrm{loss} = \sigma^\mathrm{univ} / 4(1+\tan^2\lambda)$ \\ See Eq.~\eqref{eq:mayle_in}.\vspace{4pt} \end{tabular}  & \\
\hspace{20pt} Slow short-range loss & $\bar\Gamma \not\gg \bar\gamma C^{-2}$ & \begin{tabular}{@{}l@{}}Loss cross section suppressed by $\frac{\bar\Gamma}{\bar\gamma C^{-2}+\bar\Gamma}$ \\ See Eq.~\eqref{eq:mayle_in}.\vspace{4pt}\end{tabular} & \\
\hspace{10pt} Overlapping resonances & $\rho\bar\Gamma \gg 1$ & Nonuniversal loss with $y=1/4$, see Eq.~\eqref{eq:quarter}. & Energy independent for $E < \bar\Gamma$. \\
\hline\hline
\end{tabular}
\end{center}
\end{table*}

\section{Quantum defect theory}

Collisions between ultracold molecules are described by quantum defect theory (QDT)\cite{mies:84a,mies:84b,burke:98}.
Here, one separates the short-range and long-range physics by replacing the full multi-channel scattering problem by single-channel scattering on a reference potential that asymptotically approaches the physical long-range potential.
The short-range physics can then be accounted for as a boundary condition in the long-range scattering problem on the simplified reference potential.
This boundary condition is usually taken to be energy independent for ultracold collisions,
where the collision energy $E$ and the channel wavenumber $k=\hbar^{-1}\sqrt{2\mu E}$ approach zero,
where $\mu$ is the reduced mass.
The long-range physics is described by QDT parameters $C^{-2}$, $\tan\lambda$, and $\tan\xi$.
The parameter $C^{-2}$
represents the short-range amplitude of the energy-normalized regular solution,
\emph{i.e.},\ it describes how the long-range interaction limits the molecules reaching short-range.
The parameter $\xi$ represents the scattering phase of the regular solution for the reference potential,
whereas $\lambda$ describes the phase difference between the short-range and long-range normalizations.
For simple reference potentials, these QDT parameters are known exactly.
This includes the van der Waals (vdW) potential, $-C_6 R^{-6}$, for which the low-energy $s$-wave behavior is
\begin{align}
C^{-2}(E,L=0) &= k\bar{a} \left[1 + \left(1-a/\bar{a}\right)^2\right], \nonumber \\
\tan\lambda(E,L=0) &= 1-a/\bar{a}, \nonumber \\
\tan\xi(E,L=0) &= -k a,
\end{align}
where $a$ is the scattering length and $\bar{a} = 2\pi/\Gamma(\frac{1}{4})^2 (2\mu C_6/\hbar)^{1/4}$ is the mean scattering length.
This permits computing the scattering matrix for a single {open} channel as
\begin{align}
S^\mathrm{phys} &= \exp(2 i \xi) \frac{1+i C^{-2} [(K^{\mathrm{SR}})^{-1} - \tan\lambda]^{-1}}{1-i C^{-2} [(K^{\mathrm{SR}})^{-1} - \tan\lambda]^{-1}},
\label{eq:Sphys}
\end{align}
where $K^{\mathrm{SR}}$ represents the \emph{resonant} contribution to the short-range reactance matrix~\cite{mitchell:10},
and a background phase shift is described completely by QDT.

After obtaining the physical scattering matrix,
one can calculate observable elastic and loss cross sections as
\begin{align}
\sigma^{\mathrm{el}} &= \Delta \frac{\pi \hbar^2}{2\mu E} |1-S^\mathrm{phys}|^2, \nonumber \\
\sigma^{\mathrm{loss}} &= \Delta \frac{\pi \hbar^2}{2\mu E} \left(1-\left|S^\mathrm{phys}\right|^2 \right),
\label{eq:xsec}
\end{align}
and corresponding loss rates as $k_2=v\sigma$, where $v$ is the velocity and $\Delta=2$ for indistinguishable particles.

An important special case of the short-range boundary condition is universal loss \cite{idziaszek:10}, where all molecules that reach short range are lost.
In this case, $S^{\mathrm{SR}} = 0$, and correspondingly $K^{\mathrm{SR}} = i (1-S^{\mathrm{SR}})/(1+S^{\mathrm{SR}}) = i$.
The loss rate can be computed from the sub-unitarity of the physical $S$ matrix of Eq.~\eqref{eq:Sphys} which yields
\begin{align}
1-|S^\mathrm{phys}|^2 &= \frac{4 C^{-2}}{(1+C^{-2})^2+\tan^2\lambda} = 4 k \bar{a},
\label{eq:Suniv}
\end{align}
where the second step applies to ultracold $s$-wave collisions for vdW potentials only.
As all molecules that reach short range are lost, the loss rate becomes independent of the short-range phase, $a/\bar{a}$,
and is hence referred to as universal.
The short-range loss is often expressed as a parameter $y = (1-|S^{\mathrm{SR}}|)/(1+|S^{\mathrm{SR}}|)$,
between 0 and 1,
where $y=0$ corresponds to no loss, and $y=1$ corresponds to universal loss, respectively.
For incomplete short-range loss, the physical loss rate can be larger or smaller than the universal rate.
For small $y$ the physical loss scales as $y C^{-2}$,
where $C^{-2}$ can be interpreted as a probability that the molecules traverse the long-range potential \cite{idziaszek:10}.

\section{Lossless resonance states}

To model the high density of states of molecule-molecule collision complexes,
Mayle \emph{et al.}\cite{mayle:12,mayle:13} considered the short-range reactance matrix to be dominated by a dense set of resonances drawn from a statistical distribution
\begin{align}
K^{\mathrm{SR}} = - \frac{1}{2} \sum_\nu \frac{\gamma_\nu}{E-E_\nu}.
	\label{eq:Ksr}
\end{align}
The resonance positions are distributed according to the Wigner-Dyson distribution with mean spacing $\rho^{-1}$,
which is appropriate for the classically chaotic dynamics that is thought to result from the collision complex' high density of states.
We note that Eq.~\eqref{eq:Ksr} describes the \emph{resonant} contribution to the short-range reactance matrix~\cite{mitchell:10},
whereas a background phase shift is described completely by QDT.
Unlike typical applications of QDT, the statistical short-range reactance matrix, $K^{\mathrm{SR}}$, is highly energy dependent.
There is little hope to accurately compute $K^{\mathrm{SR}}$, but the density of short-range resonances can be computed\cite{christianen:19b} and this may suffice to make statistical predictions.

Mayle \emph{et al.}\cite{mayle:12} then compute the elastic cross section for scattering due to these resonances
assuming there are many isolated resonances in the energy range we wish to average cross sections over.
Hence, it suffices to consider an averaged\footnote{i.e., integrated and multiplied by the density of resonances, $\rho$} cross section due to a single representative resonance,
replacing the individual resonance widths $\gamma_\nu$ by their mean $\bar\gamma$,
\begin{align}
&\left\langle |1-S^\mathrm{phys}|^2 \right\rangle \nonumber \\
&= \rho \int dE \frac{\bar\gamma^2 C^{-4}}{(E-E_0+\bar\gamma\tan\lambda)^2 + (\bar\gamma/2)^2 C^{-4}} \nonumber \\
&= 2\pi \rho \bar\gamma C^{-2}.
\label{eq:mayle_el}
\end{align}
Assuming $\bar\gamma=2/\pi\rho$, Mayle \emph{et al.}\cite{mayle:13} claim that this agrees exactly with the \emph{loss} cross section due to universal loss,
and postulate that this represents the cross section for loss by complex formation, although what is computed is really an \emph{elastic} cross section.

It is worth noting that the Mayle cross section is on the order of the universal cross section but, unlike claimed in Ref.~\cite{mayle:13}, not equal to it,
compare Eq.~\eqref{eq:Suniv} and~\eqref{eq:mayle_el} for $\bar\gamma=2/\pi\rho$.
The Mayle cross section is missing the denominator $1+\tan^2\lambda$.
This can be understood as the effect of $\tan\lambda$ is only to shift resonance positions,
which does not affect the cross section averaged over many resonances.
This means the Mayle cross section has an additional factor $1+(a/\bar{a}-1)^2$ which represents a dependence on the short-range phase that universal loss lacks.
Furthermore, the physically motivated mean coupling that describes classically chaotic dynamics is the Weisskopf estimate $\bar\gamma=1/2\pi\rho$.
This is a factor of four smaller than what was assumed by Mayle \emph{et al.}\cite{mayle:12,mayle:13}.
As a result, if the Weisskopf estimate holds, the cross section should be a factor of four smaller than universal loss,
and have an additional dependence on the short-range phase shift that the universal loss rate does not have.

\section{Non-overlapping lossy resonances}

Here, we extend the short-range reactance matrix to describe the finite lifetime of the resonance states.
The microscopic origin of this loss of collision complexes could be photo-excitation,
collisions of the complex with a third molecule leading to three-body recombination,
chemical reactions,
or simply the inability to trap collision complexes.
This loss is accounted for by an imaginary component of the resonance energy $E_\nu \rightarrow E_\nu - i \Gamma_\nu/2$, leading to
\begin{align}
K^{\mathrm{SR}} = - \sum_\nu \frac{\gamma_\nu/2}{E-E_\nu+i\Gamma_\nu/2}.
\label{eq:lossyK}
\end{align}
We stress this intuitive result rigorously describes resonances with elastic partial widths, $\gamma_\nu$, which are not modified,
in the presence of inelastic loss with partial widths, $\Gamma_\nu$, which broadens the resonances to a total width $\gamma_\nu+\Gamma_\nu$ \cite{landau:59,messiah:69,mitchell:10}.

Next, we compute the loss cross section from the sub-unitarity of the resulting $S$-matrix, Eq.~\eqref{eq:Sphys},
\begin{align}
1-|S^\mathrm{phys}|^2 &= \frac{4 C^{-2} \Im(K^{\mathrm{SR}})}{1+2C^{-2} \Im(K^{\mathrm{SR}}) + C^{-4} |K^{\mathrm{SR}}|^2},
\end{align}
where we have dropped $\tan\lambda$, which only shifts the position of an isolated resonance and hence does not affect a thermal average over many resonances.
Again, for a single representative resonance we have
\begin{align}
|K^{\mathrm{SR}}|^2 &= \frac{\bar\gamma^2}{4(E-E_0)^2+\bar\Gamma^2}, \nonumber \\
\Im(K^{\mathrm{SR}}) &= \frac{\bar\gamma \bar\Gamma}{4(E-E_0)^2+\bar\Gamma^2}.
\end{align}
Hence we obtain
\begin{align}
\left\langle 1-|S^\mathrm{phys}|^2 \right\rangle &= 2\pi \rho \bar\gamma C^{-2} \frac{\bar\Gamma}{\bar\gamma C^{-2} +\bar\Gamma},
\label{eq:mayle_in}
\end{align}
which, apart from the factor $\bar\Gamma/[\bar\gamma C^{-2}+\bar\Gamma]$, agrees exactly
\footnote{{%
  \unexpanded{\makeatletter\let\@bibitemShut\relax\makeatother}%
In Eq.~\eqref{eq:mayle_in}, replacing the resonance widths by the mean widths is not quite correct as $\langle f(\gamma)\rangle_\gamma \neq f(\langle \gamma\rangle_\gamma)$.
In Eq.~\eqref{eq:mayle_el}, this is not a problem as the elastic cross section integrated over a resonance depends linearly on the resonance width, which is replaced by the mean width upon averaging over many resonances.
By contrast, Eq.~\eqref{eq:mayle_in} is not linear in $\gamma$ or $\Gamma$.}}
with $\left\langle |1-S^\mathrm{phys}|^2 \right\rangle$ in the absence of loss, Eq.~\eqref{eq:mayle_el}.
Hence, for $\bar\Gamma\gg \bar\gamma C^{-2}$ the inelastic cross section agrees exactly with the elastic cross section in the absence of loss, considered by Mayle \emph{et al.}\cite{mayle:12}.
If short-range loss is slow, the additional factor suppresses loss to below the Mayle rate.

Interestingly, the condition for recovering the Mayle result is $\bar\Gamma\gg \bar\gamma C^{-2}$, not just $\bar\Gamma\gg \bar\gamma$.
This latter condition states that the destructive loss of collision complexes is fast compared to the dissociation of the short-range complex, which occurs at rate $\bar\gamma$.
Instead, there is a dependence on the QDT parameter $C^{-2}$ that describes transmission through the long-range potential.
This extends the effective lifetime of the collision complex if the long-range potential impedes dissociation,
{\emph{i.e.},\ this naturally describes the ``re-collisions'' discussed in Ref.~\cite{bause:21}.
For a vdW potential and $a=\bar{a}$, for $s$-wave collisions $C^{-2} = k\bar{a}$, which could be considerably smaller than 1 depending on the temperature.
For $p$-wave collisions $C^{-2} \approx 1.064 k^3 \bar{a}^3$, which would be smaller still at ultracold temperatures due to suppressed tunneling through the $p$-wave centrifugal barrier.
We note that the extension of the effective lifetime by limited transmission through the long-range potential was previously not considered in the rate equations governing the loss of molecules through sticky collisions \cite{mayle:13,gregory:20}.

\section{Overlapping lossy resonances}

Here, we consider the case where the loss is so fast that $\bar\Gamma\gg 1/\rho$ such that the resonances start to overlap.
This puts us in the regime of Ericson fluctuations of the cross section \cite{mitchell:10,ericson:60}.
These fluctuations are on the order of the cross section itself,
and lead to correlations between the cross section at different energies on the scale of $\bar\Gamma$.
If $\bar\Gamma \gg k_BT$, these fluctuations occur on energy scales that are not sampled thermally,
such that the resonant short-range wavefunction will appear independent of energy.
The reason for this is that many overlapping resonances -- much wider than $k_BT$ -- contribute at a single collision energy.
This means a statistical treatment may be applied even if the density of resonances is low compared to the energy window over which cross sections are sampled.
This is the case for many ultracold molecules \cite{christianen:19b}.
We note that unlike what was assumed in the discussion Ref.~\cite{mayle:12},
losses increase the resonance widths without affecting the \emph{elastic} partial width, $\bar\gamma$.

For many overlapping resonances, the sum over resonances that contribute to the reactance matrix, Eq.~\eqref{eq:lossyK}, can be approximated as an integral yielding
\begin{align}
K^\mathrm{SR} &= - \rho\int dE_\nu \frac{\gamma_\nu/2}{E-E_\nu + i \Gamma_\nu/2} \nonumber \\
&= i \pi \rho \bar\gamma/2,
\label{eq:quarter}
\end{align}
where in the second step we have assumed each of the many contributing resonances can be replaced by a representative mean resonance with $\gamma_\nu=\bar\gamma$ and $\Gamma_\nu=\bar\Gamma$.
We note that this represents a coherent average of many resonances at a single, well-defined energy,
which should be contrasted with the incoherent classical energy average of isolated resonances, considered above.
Using the Weisskopf estimate $\bar\gamma=(2\pi\rho)^{-1}$ this yields $K^\mathrm{SR}=i/4$,
corresponding to a QDT loss parameter $y=1/4$.

\begin{figure}
\begin{center}
\includegraphics[width=\figwidth]{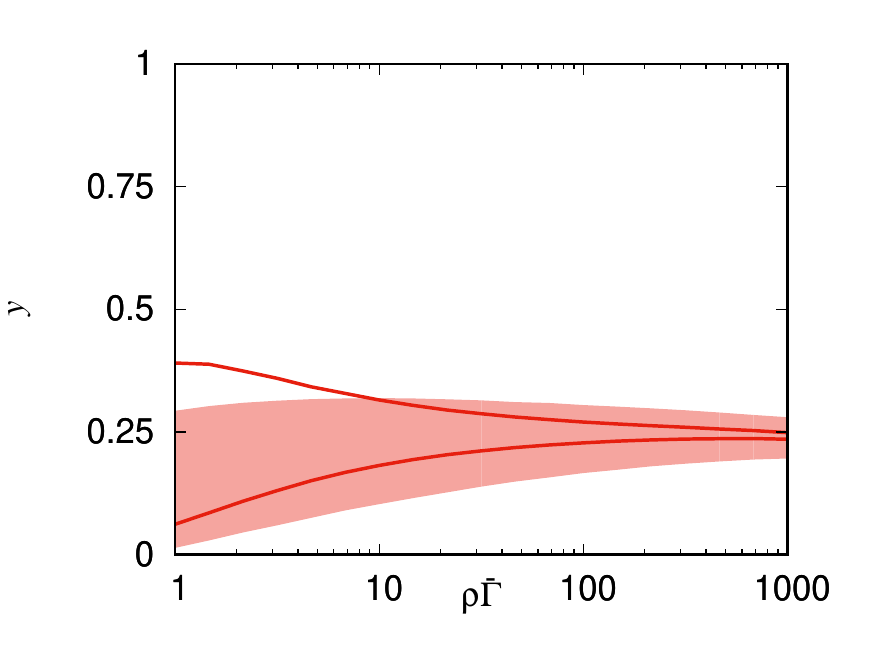}
\caption{ \label{fig:job5}
{\bf Statistically sampled loss parameters.}
Distribution of short-range loss parameters $y$ obtained by sampling realizations of statistical short-range $S$-matrices as a function of short-range loss width, $\rho\bar\Gamma$.
The red shaded area indicates the distribution for inelastic widths $\Gamma$ distributed as the square of zero-mean Gaussian-distributed couplings,
whereas the solid lines correspond to a sharply defined $\Gamma$, as appropriate in the presence of limitingly many loss channels.
Distributions are indicated by their $\pm 1\sigma$ statistical uncertainty.
}
\end{center}
\end{figure}

Equation~\eqref{eq:quarter} has been obtained by averaging over a homogeneous distribution of resonances with identical widths,
which gives an indication of the mean behavior but does not capture possible fluctuations.
We study this numerically by drawing short-range reactance matrices, Eq.~\eqref{eq:lossyK}, from the appropriate distributions.
The resonance positions are drawn from a Wigner-Dyson distribution,
whereas the half-widths are the squares of Gaussian-distributed coupling matrix elements with zero mean and variance such that $\bar\gamma=(2\pi\rho)^{-1}$ and $\bar\Gamma$ is varied.
We then numerically compute the short-range $S$ matrix and QDT short-range loss parameter $y$.
This process is repeated for many realizations of the statistical resonances.
The resulting $\pm 1\sigma$ distribution of $y$ are shown as the shaded red area in Fig.~\ref{fig:job5} as a function of $\rho\bar\Gamma$.
For high $\rho\bar\Gamma$, the distribution is closely centered around $y=1/4$,
whereas for low $\rho\bar\Gamma$, the distribution widens as few resonances contribute and the precise value of $y$ depends on their precise positions and widths.

In what is described above, both $\gamma$ and $\Gamma$ are distributed as the squares of Gaussian-distributed coupling matrix elements between the resonant state and the elastic or inelastic channel, respectively.
However, if there are many inelastic loss channels, with individual couplings distributed according to random matrix theory,
the total inelastic width $\Gamma$ has contributions from each of these and its distribution may be much narrower.
In the limiting case that each resonance has a sharply defined inelastic width, $\bar\Gamma$,
we obtain a tighter distribution of loss parameters indicated by the solid lines in Fig.~\ref{fig:job5}.
For large widths, eliminating the fluctuations in the individual inelastic widths substantially reduces the spread in the distribution of loss parameters, $y$,
whereas for small widths the distribution is similarly broad due to fluctuations in the resonance spacing and elastic widths.

Collision complexes consisting of two ultracold bialkali molecules have lifetimes that depend strongly on the masses of atomic constituents,
ranging from the microsecond scale for the lightest species involving Li atoms,
to 250~$\mu$s for the heaviest polar bialkali, RbCs \cite{christianen:19b}.
Rates of excitation of collision complexes by an optical dipole trap are typically in the $\mu$s$^{-1}$ range \cite{christianen:19a,liu:20,gregory:20}.
Hence, even for the lightest molecules, $\rho\bar\Gamma$ may be of order unity,
for which Fig.~\ref{fig:job5} shows substantial but nonuniversal short-range loss.
For the heaviest bialkali molecule, RbCs, the lifetime with respect to laser excitation is three orders of magnitude shorter than the lifetime with respect to dissociation of the complex,
which would put $\rho\bar\Gamma>1000$ and hence the present theory using the Weisskopf estimate of the mean width yields $y$ close to $1/4$, consistent with the experimental result $y=0.26(3)$ \cite{gregory:19}.

\section{Thermal averaging}

\begin{figure}
\begin{center}
\includegraphics[width=\figwidth]{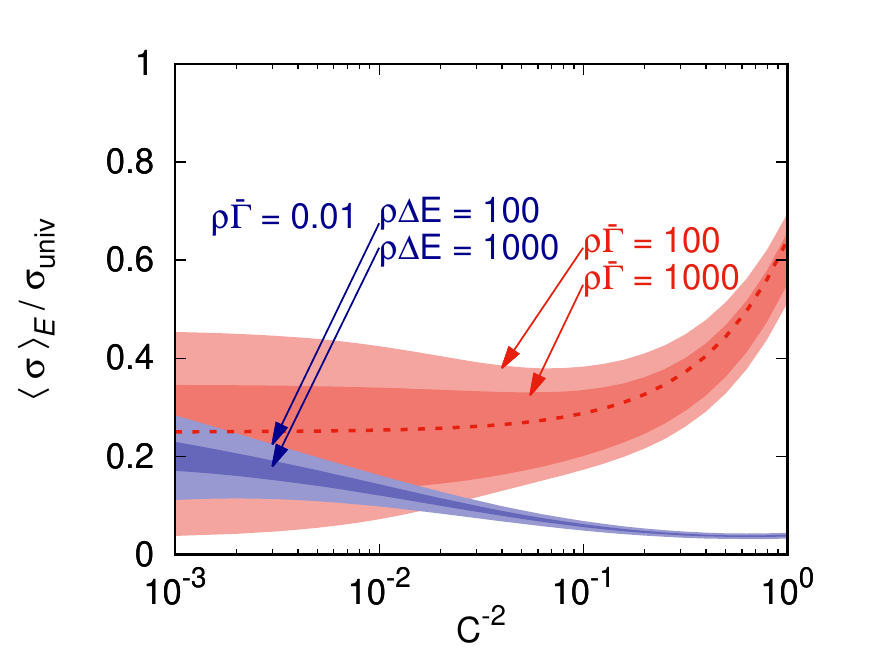}
\caption{ \label{fig:job7}
{\bf Statistically sampled energy-averaged loss cross sections,}
normalized to universal loss for $a=\bar{a}$ as a function of the wavefunction amplitude at short-range, $C^{-2}$.
Shaded areas indicate $1\sigma$ statistical uncertainty.
For slow short-range loss and $\bar\gamma=1/2\pi\rho$, shown in blue,
the loss rate approaches the Mayle result of one fourth of the universal rate only at small $C^{-2}$ where $\bar\gamma C^{-2} \ll \bar\Gamma$.
For fast loss $\rho\bar\Gamma\gg 1$, the loss becomes independent of energy and shows a large statistical uncertainty characteristic of Ericson fluctuations,
indicated as the light and dark red shaded areas corresponding to $\rho\bar\Gamma=100$ and $1000$, respectively.
}
\end{center}
\end{figure}

Next, we consider observable loss cross sections or rates,
which depend not only on the short-range reactance matrix, $K^\mathrm{SR}$, but also on the transmission through the long-range potential, $C^{-2}$, and a thermal average.
A complication is that the short-range density of states and the long-range interactions occur on independent characteristic energy scales.
We first consider a simplified discussion where we take $C^{-2}$ as an energy-independent parameter,
and energy-average only the statistical short-range reactance matrix.
Rather than using the Maxwell-Boltzmann distribution,
we use a square energy window of width $\Delta E$ to average the sub-unitarity of the scattering matrix as 
\begin{align}
{\langle1-|S^\mathrm{phys}|^2\rangle_E} = {\frac{1}{\Delta E} \int_0^{\Delta E} 1-|S^\mathrm{phys}(d)|^2 \mathrm{d}E},
\end{align}
yielding an energy-averaged loss cross section.

The resulting energy-averaged loss cross section normalized to the universal cross section is shown in Fig.~\ref{fig:job7}.
In blue, for non-overlapping resonances with weak loss $\rho\bar{\Gamma}=0.01$, we observe that for small $C^{-2}$ the loss cross section approaches the Mayle result,
which is a factor of four below the universal result for the Weisskopf width $\bar\gamma=1/2\pi\rho$ and $a=\bar{a}$, used here.
This is approached for small $C^{-2}$ such that $\bar\Gamma\gg \bar\gamma C^{-2}$ despite the small loss rate,
whereas the loss cross section is suppressed further for larger $C^{-2}$, where short-range loss is slower than dissociation,
see Eq.~\eqref{eq:mayle_in}.
Hence, for slow short-range loss,
the loss cross section can be suppressed below the universal rate by increasing the transmission, $C^{-2}$, by raising the temperature or inducing longer ranged interactions.
The statistical uncertainty is reduced if more resonances are sampled thermally, as can be seen from the light and dark shaded areas corresponding to $\rho\Delta E=100$ and $1000$, respectively.
In red, results for rapid loss with $\rho\bar{\Gamma}=100$ and $1000$ are shown as the light and dark shaded areas.
The variation is large as expected in the Ericson regime where fluctuations of the cross section are of the same order as the cross section itself \cite{mitchell:10,ericson:60}.
The mean cross section, \emph{i.e.}, assuming $K^\mathrm{SR}=i/4$, is shown as the dotted line.
We confirm numerically that the loss is independent of energy for $\Delta E \ll \bar\Gamma$.

For overlapping resonances, $\rho\bar\Gamma \gg 1$, cross sections display Ericson fluctuations \cite{mitchell:10,ericson:60}.
This is illustrated in Fig.~\ref{fig:ericson},
which shows the autocorrelation function of the cross section at different energies, $E$ and $E+\Delta E$.
Cross sections computed here account only for the short-range physics, \emph{i.e.},\ assuming $C^{-2}=1$,
and do not account for threshold effects present that may give rise to an additional energy dependence in physical ultracold collisions, for example for ultracold $p$-wave collisions.
Dashed lines show numerical results whereas the solid lines show the expected Lorentzian behavior, $1/(1+(\Delta E/\bar\Gamma)^2)$,
where the width of the distribution is set by the sharply defined inelastic loss width, $\bar\Gamma$.
Thus, for overlapping resonances the cross sections will be constant over an energy range, $\bar\Gamma$, much wider than the mean spacing between resonances, $1/\rho$,
such that cross sections are independent of whether or not the collision energy coincides with a resonance.
This underscores the present statistical treatment is predictive even where the resonances are sparse compared to $k_BT$, the energy range sampled thermally.

\begin{figure}
\begin{center}
\includegraphics[width=\figwidth]{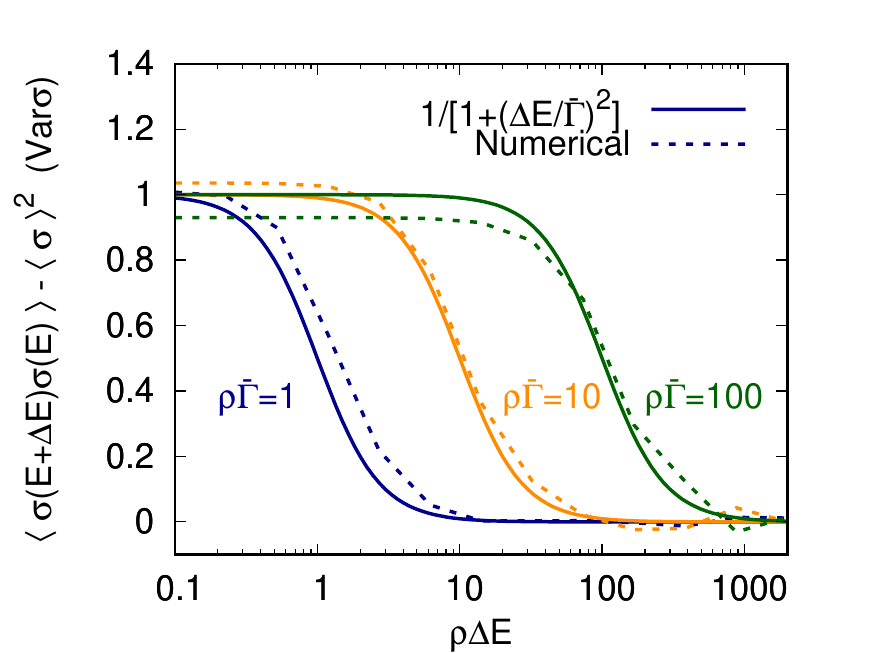}
	\caption{ \label{fig:ericson}
	{\bf Ericson fluctuations.}
	Autocorrelation function of the loss cross section is Lorentzian on an energy scale $\bar\Gamma$, characteristic of Ericson fluctuations \cite{mitchell:10,ericson:60}.
	This shows cross sections fluctuate over an energy scale set by $\bar\Gamma$, which may be considerably wider than the spacing between individual resonances, $\rho^{-1}$,
	underscoring the present statistical treatment is predictive even where the resonances are sparse compared to $k_BT$, the energy range sampled thermally.
}
\end{center}
\end{figure}

After this initial simplified discussion,
we consider the proper thermal average of the loss rate over a Maxwell-Boltzmann distribution at finite temperature for (a) bosonic RbCs ($\rho=5.3~\mu$K$^{-1}$), (b) bosonic NaK ($\rho=0.12~\mu$K$^{-1}$), and (c) fermionic NaK molecules ($\rho=0.37~\mu$K$^{-1}$),
for which loss rates as a function of temperature are shown in Fig.~\ref{fig:real}.
In each case, temperature relative to the density of states of the collision complex, $\rho k_BT$, determines the number of statistical resonances sampled thermally,
while temperature relative to the vdW energy, $k_BT/ E_6$, determines the short-range amplitude $C^{-2}$.
Here, the vdW energy scale is $E_6 = \hbar^3/ (8\mu^3 C_6)^{1/2}$.
Accounting for the energy dependence of $C^{-2}$ increases the overall variance in the loss rates, indicated by the $\pm1~\sigma$ shaded areas.
Secondly, low $C^{-2}$ occur only for low temperatures where one does not thermally average over many resonances.
For isolated resonances with $\rho\bar\Gamma=0.01$ shown in blue,
this leads to essentially complete uncertainty in the loss rate at low temperatures and equivalently small $C^{-2}$.
This implies that in practice ultracold molecules do not realize the scenario originally envisioned by Mayle \emph{et al.}\ where one thermally averages over many isolated resonances.
However, the mean loss rate averaged over many realizations (solid blue line) follows Eq.~\eqref{eq:mayle_in} (dashed blue line) reasonably well.
Loss rates for overlapping resonances are shown in orange and red for $\rho\bar\Gamma=100$ and $1000$, respectively.
In this case, the observable loss rate is well defined at all temperatures as even at the lowest temperatures one samples many resonances due to their overlap,
although also in this case there exists a substantial spread in the possible loss rates.
Shown in pink are loss rates for sharply defined $\Gamma=1000/\rho$,
which results in a sharply defined loss rate described by $y=1/4$.
Finally,
the loss rates of RbCs and NaK molecules show large similarities even though their density of states differ by a factor 40,
suggesting that these results can be considered somewhat universal.

\begin{figure}
\begin{center}
\includegraphics[width=\figwidth]{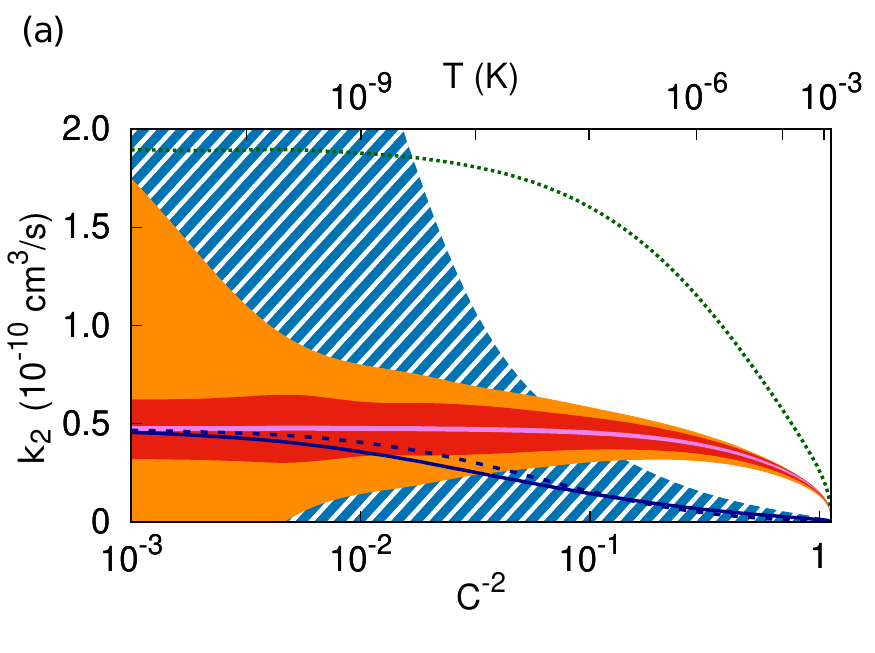}
\includegraphics[width=\figwidth]{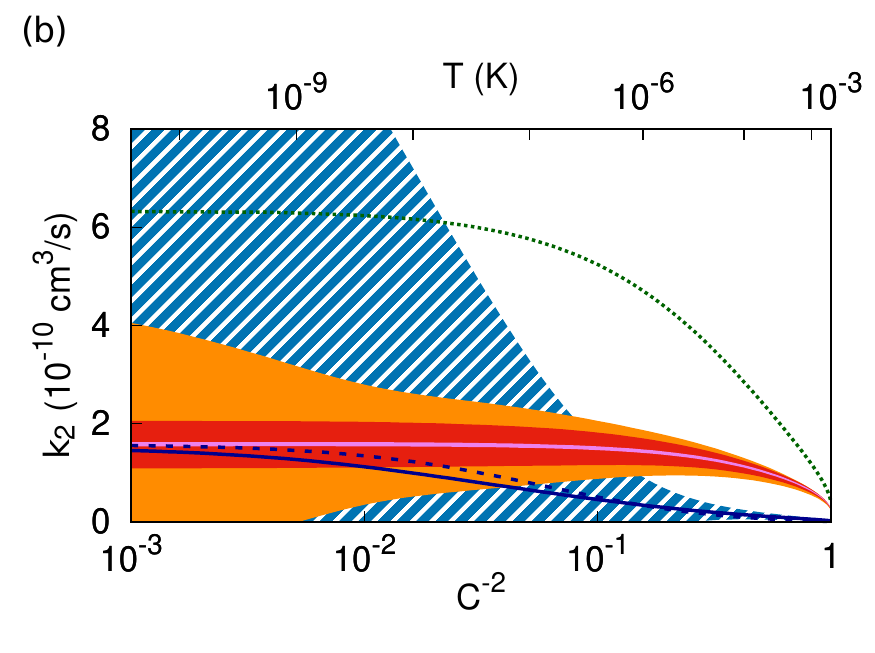}
\includegraphics[width=\figwidth]{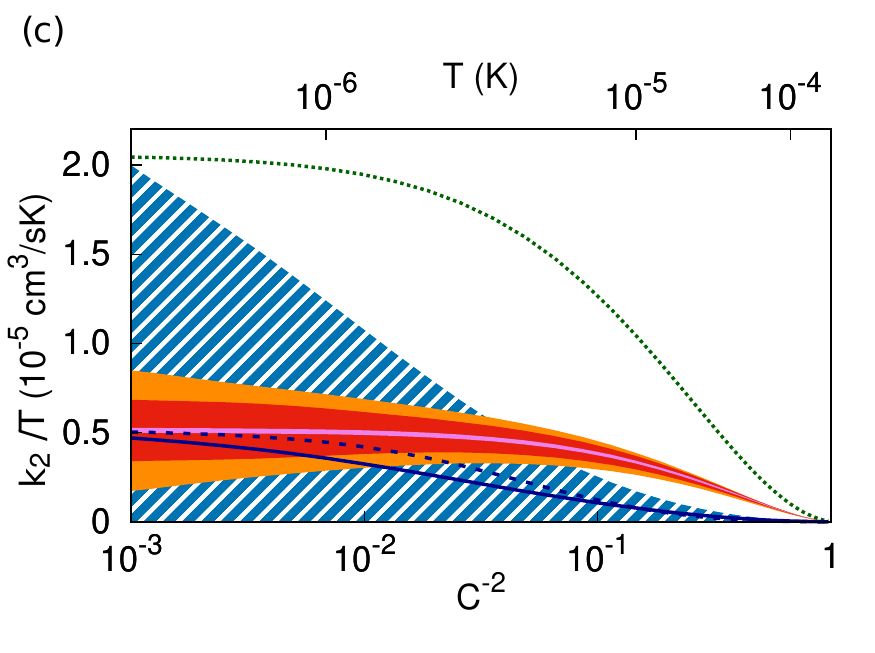}
\caption{ \label{fig:real} 
{\bf Statistically sampled thermally averaged loss rates,}
as a function of temperature or equivalently $C^{-2}$ for $a=\bar{a}$.
The three panels show (a) bosonic RbCs, (b) bosonic NaK, and (c) fermionic NaK.
Note the change in temperature scale due to $p$-wave collisions for fermionic molecules.
Green dotted lines indicate the universal loss rate,
blue solid lines indicate the mean loss rate for isolated resonances with $\rho\bar\Gamma=0.01$,
which reasonably follows $\bar\Gamma/[\bar\gamma C^{-2}+\bar\Gamma]$ indicated as the blue dashed line, see Eq.~\eqref{eq:mayle_in}.
Shaded areas indicate $1\sigma$ statistical uncertainty.
The striped blue area indicates isolated resonances with $\rho\bar\Gamma=0.01$,
orange overlapping resonances with $\rho\bar\Gamma=100$,
red $\rho\bar\Gamma=1000$, 
and pink indicates sharply defined $\Gamma=1000/\rho$ (remaining three shaded areas, outside in).
}
\end{center}
\end{figure}

\section{Nonuniversality}

A qualitative conclusion to be drawn from the present work is that we should expect \emph{nonuniversal} loss even if short-range loss is orders of magnitude faster than the lifetime of collision complexes.
In this case, one would have naively expected complete loss of short-range complexes, $y=1$, and hence universal scattering.
Figure~\ref{fig:nonuniversal} shows the dependence of nonuniversal cross sections on an unknown background scattering length $a$,
whereas all results shown above used $a=\bar{a}$ as an example.

\begin{figure}
\begin{center}
\includegraphics[width=\figwidth]{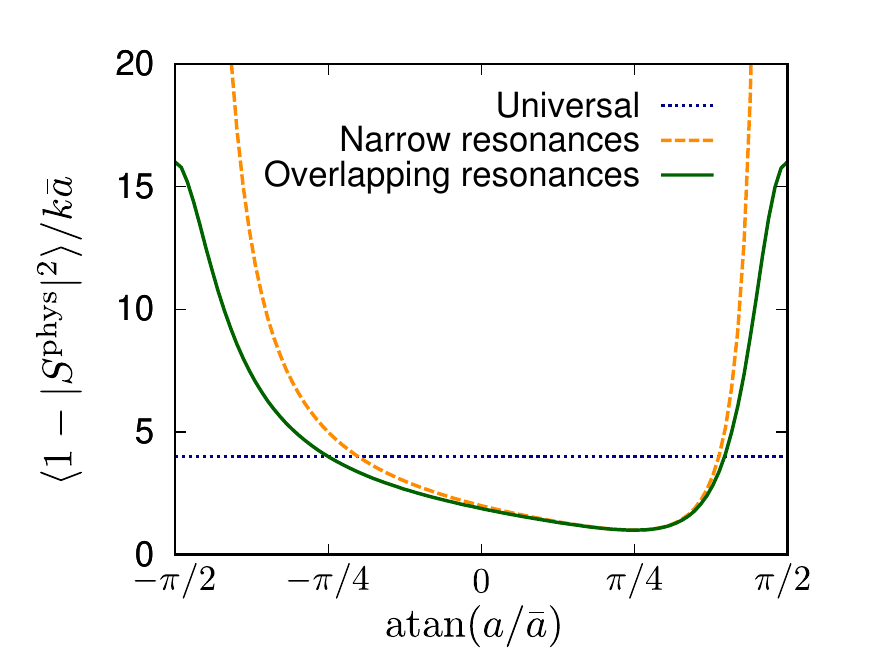}
        \caption{ \label{fig:nonuniversal}
        {\bf Nonuniversal cross section versus background phase shift}, parameterized by $\atan(a/\bar{a})$.
	Quantum defect parameters correspond to $s$-wave collisions on a van der Waals interaction potential.
	Universal loss yields $1-|S^\mathrm{phys}|^2 = 4 k\bar{a}$, see Eq.~\eqref{eq:Suniv}.
	For lossy but narrow isolated resonances, ${\langle1-|S^\mathrm{phys}|^2\rangle} = k\bar{a} [1+(1-a/\bar{a})^2]$, see Eq.~\eqref{eq:mayle_in}.
	For overlapping lossy resonances, $K^\mathrm{SR}=i/4$, which yields $1-|S^\mathrm{phys}|^2 = k\bar{a} 16 [1+(1-a/\bar{a})^2]/[16+(1-a/\bar{a})^2]$.
}
\end{center}
\end{figure}

To better understand the origin of this nonuniversality, we first consider a resonant collision ${E=E_\nu}$ for an isolated resonance $\nu$,
such that the short-range reactance matrix ${K^\mathrm{SR} = i \gamma/\Gamma}$.
The short-range wave function $f^\mathrm{SR} - g^\mathrm{SR} K^\mathrm{SR}$,
where WKB approximations to the regular and irregular solutions are
\begin{align}
f^\mathrm{SR}&=k^{-1/2} \sin[\int^R k(R')~dR'],\nonumber \\
g^\mathrm{SR}&=-k^{-1/2} \cos[\int^R k(R')~dR'],
\end{align}
and $k(R)$ is the local wavenumber.
For $K^\mathrm{SR}=0$ -- by definition -- the short-range wavefunction is given by the regular short-range solution.
For elastic scattering on resonance, $|K^\mathrm{SR}|\rightarrow \infty$ and the short-range wavefunction approaches the irregular solution.
For lossy scattering on resonance, $K^\mathrm{SR}= i \gamma/\Gamma$,
such that for $\gamma\gg \Gamma$ we recover elastic scattering as expected.
The lifetime due to elastic decay of the resonance is short compared to the time scale for short-range loss, leading to a small loss rate.
In the other extreme, for very fast short-range loss, $\gamma \ll \Gamma$, the loss on resonance also becomes very small.
At fixed $\Gamma$, this can be understood as small $\gamma$ provides ineffective coupling to the short-range resonances.
At fixed $\gamma$, this result seems less intuitive as increased short-range loss rate, $\Gamma$, counter-intuitively leads to a smaller overall loss rate,
which can be interpreted as a quantum Zeno effect.
Short-range loss is maximal in the intermediate case, $\Gamma=\gamma$, where we obtain $K^\mathrm{SR}= i$ such that the short-range wavefunction describes a plane wave approaching the origin,
with no flux returning from the origin,
\emph{i.e.},\ it models complete universal loss at short range.
In this case, there is both an effective coupling to short-range resonances, as well as effective loss of the complexes when formed.
If the short-range loss rate is increased, loss is suppressed by the quantum Zeno effect,
whereas if the short-range loss rate is decreased loss is also suppressed as complexes decay elastically before they undergo loss.

Returning to the case of multiple resonances,
in the non-overlapping case it can be understood that the loss rate increases with $\rho\bar\gamma$, the ratio of the coupling and the mean level spacing,
as this determines the fraction of collision energies that are close to resonance.
It appears from Eq.~\eqref{eq:mayle_in} that the loss rate increases with $\rho\bar\gamma$ without an upper bound,
but this result applies only to non-overlapping resonances and hence small $\rho\bar\gamma$.
In the overlapping case we consider $\bar\Gamma \gg \bar\gamma$, such that the contribution of each resonance to the reactance matrix decreases as $\bar\gamma/\bar\Gamma$.
However, $\bar\Gamma$ also broadens the resonances meaning that the number of contributing resonances increases as $\bar\Gamma \rho$,
which leads to an overall scaling $K^\mathrm{SR} \propto \rho\bar{\gamma}$.
We stress that here it is the reactance matrix, rather than the loss rate, that scales with $\rho\bar{\gamma}$ and that leads to a maximum short-range loss parameter $y=1$ for $K^\mathrm{SR}=i$.
This universal loss is obtained for $\bar{\gamma}=2/\pi\rho$,
and the loss parameter $y$ is reduced by either fast elastic scattering for larger $\bar{\gamma}$ or by quantum Zeno suppression for smaller $\bar\gamma$.
For classically chaotic dynamics, however, the physically motivated width is given by the Weisskopf estimate of random matrix theory, $\bar{\gamma}\rho=1/2\pi$, which results in nonuniversal loss with $y=1/4$.

\section{Comparison to Experiment}

In order to compare the present theory to experimentally measured loss rates of ultracold molecules we need to account for the spread in the possible theoretical loss rates that is caused by two factors.
First, there is the statistical spread caused by uncertainty in the position and widths of the contributing short-range resonances, see Fig.~\ref{fig:real}.
Second, the dependence on the background phase shift can result in cross sections that happen to be close to the universal cross section even if the short-range loss is nonuniversal, see Fig.~\ref{fig:nonuniversal}.
Deviations that are not explained by these factors must reflect the breakdown of the basic assumption of this model,
namely that the short-range resonance states are described by classically chaotic dynamics.
We note that, similar to what is discussed by Croft \emph{et al.}~\cite{croft:20},
the present theory predicts loss rates dependent on the ratio of the mean elastic coupling to the mean level spacing, $\rho\bar\gamma$.
Thus, deviations of experimentally determined loss parameters from the statistical spread described here could be interpreted as a deviation from the Weisskopf width.
Rather than modifying the mean elastic width of the entire set of short-range states,
it is also possible that the dynamics is determined by a small number of non-chaotic states.

For many ultracold molecules, including KRb \cite{ni:08}, NaK \cite{park:15,yan:20,bause:21,gersema:21}, NaRb \cite{ye:18,guo:18,gersema:21}, and CaF \cite{cheuk:20} molecules, observed collisional loss is close to universal.
For nonreactive molecules, universal loss was initially interpreted as evidence of sticky collisions due to chaotic dynamics,
but in light of the present work this should instead be taken as evidence of the opposite; non-chaotic dynamics.
This is surprising as chaotic dynamics has been observed in classical simulations~\cite{croft:14,klos:21}, quantum scattering calculations~\cite{croft:17,croft:17b}, and experimental reaction product distributions~\cite{liu:21}.
However, it is not completely clear whether initial observations of close-to-universal loss are quantitatively inconsistent with the presented theory.
So far, short-range loss parameters have been determined quantitatively only for RbCs molecules,
which required accurate measurements of the collisional loss rate and their temperature dependence~\cite{gregory:19}.
This yielded $y=0.26(3)$, coincidentally in close agreement with the present theoretical result, $y=1/4$.
Similar fitting to constrain $y$ and $a/\bar{a}$ for NaRb collisions does not exclude $y=1/4$ to within the experimental uncertainty~\cite{bai:19},
although the most probable value appears to be $y=0.5$ in the vibrational ground state and closer to universal in the first vibrationally excited state.

We note that the model developed here is also applicable to reactive molecules,
as long as the chemical reactions are mediated by the formation of collision complexes.
If chemical reactions can occur \emph{directly}, without formation of short-range resonance states,
losses in the nonresonant background scattering described by QDT would also need to be accounted for.
Measurements of the lifetime of reactive KRb-KRb collision complexes \cite{liu:20}, however,
agree with estimates based of the Rice-Ramsperger-Kassel-Marcus lifetime, suggesting reactions are indeed mediated by the formation of collision complexes.
Hence, our model predicts that chemically reactive losses of KRb molecules also lead to nonreactive losses described by $y=1/4$ when assuming the Weisskopf estimate of the mean coupling strength.
Experimentally, loss of fermionic KRb molecules is viewed as consistent with universal~\cite{demarco:19}, $y=1$,
although there is some variation in the reported loss rates~\cite{ospelkaus:10,demarco:19} and loss of KRb molecules in distinguishable hyperfine states has been found to be non-universal~\cite{idziaszek:10}.

As noted above, collisional loss rates close to universal can be recovered within the present framework in two ways;
either by increasing $\rho\bar\gamma$ to $2/\pi$, exactly four times the Weisskopf estimate,
or for particular background scattering lengths, see Fig.~\ref{fig:nonuniversal}.
Both of these scenarios rely on a fine tuning, either of the mean coupling strength or of the scattering length,
and we consider it unlikely this occurs for the multitude of molecules for which close-to-universal loss is observed experimentally.
Finally, in the Appendix we show for completeness that loss from non-universal collisions may appear universal without fine-tuning when averaged over the short-range phase shift.
We have no physical justification for such averaging, however, and we do not consider this an explanation for observations of universal loss of nonreactive molecules.

\section{Conclusions}

In conclusion, we have considered a simple model of scattering of ultracold molecules in the presence of a set of short-range resonances with a limited lifetime,
describing the formation and subsequent loss of classically chaotic collision complexes.
We consider various limiting cases for the short-range loss rate,
summarized in Table~\ref{table},
which range from isolated lossless resonances to strongly overlapping resonances broadened by loss.
For a high density of isolated resonances that is sampled thermally,
we verify that collisional loss cross sections are identical to the resonant elastic cross section, as postulated by Mayle \emph{et al.}\cite{mayle:12,mayle:13},
but that this cross section is nonuniversal.
Next, we considered rapid loss that broadens the short-range resonances to the point of overlapping.
Here, many overlapping resonances contribute at any collision energy,
which gives rise to energy-independent short-range loss and enables a statistical description even if the density of resonances is low compared to temperature,
which is the case for many molecules.
This leads to substantial but nonuniversal short-range losses with the short-range loss parameter approaching $y=1/4$ for extreme broadening.
The present work suggests nonuniversal loss for molecules occurs even if the loss of short-range complexes is fast compared to their lifetime, and that this occurs as a direct consequence of the chaotic dynamics of collision complexes.
Deviations of experimental loss rates from the present theory that cannot be explained by either statistical fluctuations or the dependence on the background scattering phase shift indicate non-chaotic short-range dynamics.

\section{Acknowledgements}

We are grateful to John Bohn and Kang-Kuen Ni for useful discussions.

\appendix
\section{Short-range phase-averaged loss \label{sec:appendix_averaged}}

It is interesting to note that close-to-universal loss could be observed without fine-tuning for any short-range loss parameter $y$ if only the observed loss were averaged over the background phase shift.
This is illustrated in Fig.~\ref{fig:averaged} which shows loss rates averaged over $\atan(a/\bar{a})$ in units of the universal rate for $y=0.1$, $0.25$, and $0.5$, as a function of the collision energy relative to the characteristic vdW energy scale.
In the classical regime where $E/E_6 \gg 1$, the loss rate approached is a factor $4y/(1+y)^2$ lower than the universal rate,
whereas for very low energies the loss rate approaches the universal result.
For smaller short-range loss, $y$, this occurs for ever lower collision energies.
However, we have no physical justification for a uniform average over the short-range phase, $\atan(a/\bar{a})$.
Nevertheless, it is an interesting observation that averaging the background phase shift leads to universal loss without any fine tuning.

\begin{figure}
\begin{center}
\includegraphics[width=\figwidth]{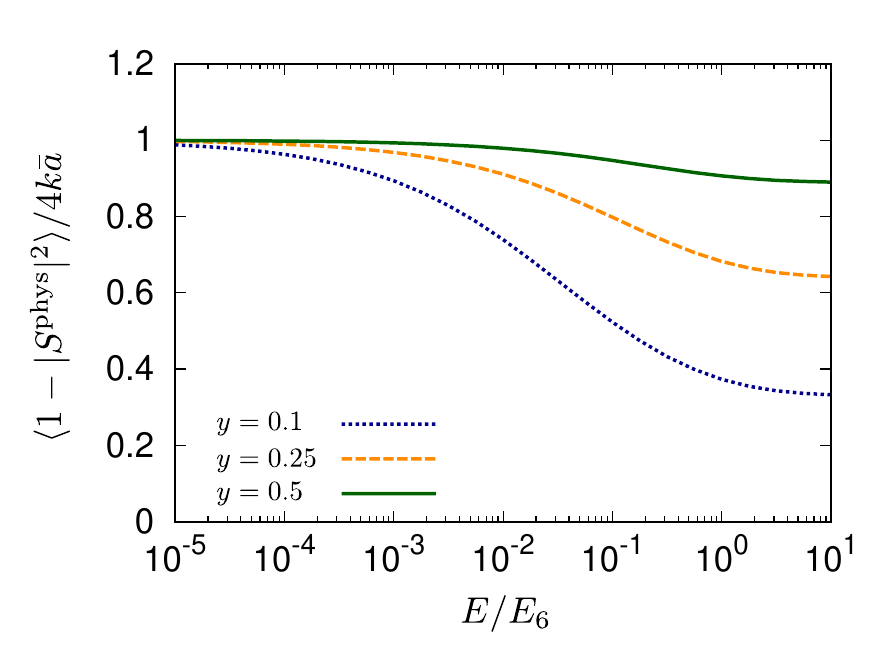}
        \caption{ \label{fig:averaged}
{\bf Loss rate averaged over the background phase shift} in units of the universal loss rate as a function of the collision energy, for $s$-wave collisions on an $R^{-6}$ potential.
In the classical regime where $E/E_6 \gg 1$, the loss rate approached is a factor $4y/(1+y)^2$ below the universal rate,
whereas for very low energies the loss rate approaches the universal result.
For smaller short-range loss, $y$, this occurs for ever lower collision energies.
}
\end{center}
\end{figure}

\end{document}